# CloudMe Forensics: A Case of Big-Data Investigation

Yee-Yang Teing, Ali Dehghantanha *Senior Member IEEE*, and Kim-Kwang Raymond Choo, *Senior Member, IEEE*

*Abstract*—The issue of increasing volume, variety and velocity of has been an area of concern in cloud forensics. The high volume of data will, at some point, become computationally exhaustive to be fully extracted and analysed in a timely manner. To cut down the size of investigation, it is important for a digital forensic practitioner to possess a well-rounded knowledge about the most relevant data artefacts from the cloud product investigating. In this paper, we seek to tackle on the residual artefacts from the use of CloudMe cloud storage service. We demonstrate the types and locations of the artefacts relating to the installation, uninstallation, log-in, log-off, and file synchronisation activities from the computer desktop and mobile clients. Findings from this research will pave the way towards the development of data mining methods for cloud-enabled big data endpoint forensics investigation.

## I. INTRODUCTION

With the advancement of broadband and pervasive media devices (e.g., smartphones and tablets), it is not uncommon to find consumer devices storage media that can hold up to Terabytes (TB) worth of data. Federal Bureau of Investigation's fifteen Regional Computer Forensic Laboratories reported that the average amount of data they processed in 2014 is 22.10 times the amount of data ten years back, up from 22TB to 5060TB [1], [2]. The increase in storage capacity had a direct impact on cloud forensic, and hence it is inevitable that big data solutions become an integral part of cloud forensics tools [3].

Due to the nature of cloud-enabled big data storage solutions, identification of the artefacts from the cloud hosting environment may be a 'finding a needle in a haystack' exercise [4]. The data could be segregated across multiple servers via virtualisation [5]. Lack of physical access to the cloud hosting environment means the examiners may need to rely on the Cloud Service Provider (CSP) for preservation of evidence at a lower level of abstraction, and this may not often be viable due to service level agreements between a CSP and its consumers [6]–[14]. Even if the location of the data could be identified, traditional practices and approaches to computer forensics investigation are unlikely to be adequate [9] i.e., the existing digital forensic practices generally require a bit-by-bit copy of an entire storage media [15]–[17] which is unrealistic and perhaps computationally infeasible on a large-scale dataset [12]. It has been demonstrated that it could take more than 9 hours to merely acquire 30GB of data from an Infrastructure as a Service (IaaS) cloud environment [18], [19] hence, the time required to acquire a significantly larger dataset could be considerably longer. These challenges are compounded in cross-jurisdictional investigations which could prohibit the transfer of evidential data due to the lack of cross-nation legislative agreements in place [7], [20]–[22]. Therefore, it is unsurprising that forensic analysis of the cloud service endpoints remains an area of research interest [22]–[29].

CloudMe (previously known as 'iCloud') is a Software as a Service (SaaS) cloud model currently owned and operated by Xcerion [30]. The CloudMe service is provided in a free version up to 19 GB (with referral program) and premium versions up to 500 GB storage for consumers and 5 TB for business users [31]. CloudMe users may share contents with each other as well as public users through email, text-messaging, Facebook and Google sharing. There are three (3) modes of sharing in CloudMe namely WebShare, WebShare+, and Collaborate. WebShare only permits one-way sharing where the recipients are not allowed to make changes to the shared folder. WebShare+ allows users to upload files/folders only, while collaborative sharing allows the recipients to add, edit or delete the content, even without the use of CloudMe client application [32]. The service can be accessed using the web User Interface (UI) as an Internet file system or the client applications, which are available for Microsoft Windows, Linux, Mac OSX, Android, iOS, Google TV, Samsung Smart TV, Western Digital TV, Windows Storage Servers, Novell's Dynamic File Services Suite, Novosoft Handy Backup etc. CloudMe is also compatible with third (3$^{rd}$) path software and Internet services, enabling file compression, encryption, document viewing, video and music streaming etc. through the web/client applications [32].

In this paper, we seek to identify, collect, preserve, and analyse residual artefacts of use CloudMe cloud storage service on a range of end-point devices. We focus on the

Yee-Yang Teing is with the Department of Computer Science, Faculty of Computer Science and Technology, Universiti Putra Malaysia, Serdang, 43400 Selangor, Malaysia and the School of Computing, Science and Engineering, University of Salford, Salford, Greater Manchester M5 4WT, UK (e-mail: teingyeeyang@gmail.com).

Ali Dehghantanha is with the School of Computing, Science and Engineering, University of Salford, Salford, Greater Manchester M5 4WT, UK (e-mail: A. Dehghantanha@salford.ac.uk).

Kim-Kwang Raymond Choo is with Department of Information Systems and Cyber Security, University of Texas at San Antonio, San Antonio, TX 78249-0631, USA (e-mail: raymond.choo@fulbrightmail.org).

residual artefacts from the client-side perspective to provide core-evidences serve as starting points for CloudMe investigation in a big data environment. We attempt to answer the following questions in this research:

1. What residual artefacts remain on the hard drive and physical memory after a user has used the CloudMe desktop client applications, and the locations of the data remnants on a Windows, Ubuntu, and Mac OS client device?
2. What residual artefacts can be recovered from the hard drive and physical memory after a user has used the CloudMe web application?
3. What Cloudme residual artefacts remain on the internal memory, and the locations of the data remnants on an Android and iOS client device?

The structure of this paper is as follows. In the next section, we describe the related work. Section III highlights the experiment environment setup. In Section IV, we discuss the traces from the storage media and physical memory dumps of the desktop clients. Section V presents the findings from mobile clients and network traffic, respectively. Finally, we conclude the paper and outline potential future research areas in Section VI.

## II. LITERATURE REVIEW

The National Institute of Standard and Technology (NIST) defines cloud computing as "*[a] model for enabling ubiquitous, convenient, on-demand network access to a shared pool of configurable computing resources (e.g., networks, servers, storage, applications, and services) that can be rapidly provisioned and released with minimal management effort or service provider interaction*" [33]. The key aspects are to provide on-demand self-service, broad network access, resource pooling, rapid edacity, and measured services. There are three cloud computing service models [33], which are Software as a Service (SaaS), Platform as a Service (PaaS), and Infrastructure as a Service (IaaS). NIST [33] also defined four deployment models as part of the cloud computing definition, which are public, private, community, and hybrid clouds. The public cloud is owned and operated by a provider organisation. Consumers can subscribe to the service for a fee, based on the storage or bandwidth usage. On the other hand, the private cloud is tailored to a single organisation's needs. The cloud infrastructure that is administered by organisations sharing common concerns (e.g., mission, security requirements, policy, and compliance considerations) are called community cloud.

Cloud computing is not without its own unique forensics challenges. Challenges such as jurisdiction differences, loss of data control, physical inaccessibility of evidences, multi-tenancy, and lack of tools for large scale distributed and virtualized systems are often cited as main causes of concern for cloud forensics [34]–[37]. Other pinpointed challenges include diverse digital media types, anonymity of IP addresses, decentralisation, and utilisation of anti-forensic and encryption techniques [36], [38], [39]. Fahdi et al. [40] found that the top three cloud forensic challenges according to digital forensic practitioners are volume of data, legal aspect, and time, while the top three challenges raised by digital forensic researchers are time, volume of data, and automation of forensic analysis.

Delving deeper into the legal challenges, Hooper et al.[20] reviewed the 2011 Australian Federal Government's Cybercrime Bill amendment on mutual legal assistance requests and concluded that laws amendment on a jurisdiction alone may not be adequate to address multi-jurisdiction investigation issues in cloud computing environments. Martini and Choo[7], Taylor et at. [41], and Daryabar et al.[10] also agree on the need for harmonious laws across jurisdictions. Simou et al. [36] and Pichan et al. [37] added that the issues of CSP dependence could exacerbate the challenges in all stages of cloud forensics (e.g., identify, preserve, analyse, and report [42], [43]). Consequently, Farina et al. [44] and Damshenas et al.[3], [11] suggested that such concerns can be mitigated through clearly-defined Service Level Agreements (SLA) between service providers and consumers.

Martini and Choo [45] proposed the first cloud forensic investigation framework, which was derived based upon the frameworks of McKemmish [46] and NIST [43]. The framework was used to investigate ownCloud[47], Amazon EC2[18], VMWare [48], and XtreemFS [49]. Quick et al.[22] further extended and validated the four-stage framework using SkyDrive, Dropbox, Google Drive, and ownCloud. Chung et al. [50] proposed a methodology for cloud investigation on Windows, Mac OSX, iOS, and Android devices. The methodology was then used to investigate Amazon S3, Google Docs, and Evernote. Scanlon et al. [51] outlined a methodology for remote acquisition of evidences from decentralised file synchronisation networks and utilised it to investigate BitTorrent Sync [52]. In another study, Teing et al. [53] proposed a methodology for investigating the newer BitTorrent Sync application (version 2.0) or any third party or Original Equipment Manufacturer (OEM) applications. Do et al. [54] proposed an adversary model for digital forensics and demonstrated how such an adversary model can be used to investigate mobile devices (e.g. Android smartwatch – Do et al. [55] and apps). Ab Rahman et al. [56], proposed a conceptual forensic-by-design framework to integrate forensics tools and best practices in development of cloud systems.

Marty [57] and Shields et al. [58] proposed a proactive application-level logging mechanism, designed to log information of forensics interest. However, Zawoad and Hasan[59] argued that the proposed solutions may not be viable in real world scenarios. Forensic researchers such as Dykstra and Sherman [60], Gebhardt and Reiser [61], Quick et al. [22], and Martini and Choo [48], on the other hand, presented methods and prototype implementations to support the (remote) collection of evidential materials using Application Programming Interfaces (API). Quick and Choo[62] and Teing et al.[25] studied the integrity of data downloaded from the web and desktop clients of Dropbox, Google Drive, Skydrive, and Symform and identified that the act of downloading files from client applications does not breach the evidence integrity (e.g., no change in the hash values), despite changes in file creation/modification time.

In addition to remote collection of evidences, scholars also studied the potential of on-device collection of cloud artefacts such as from Evernote[50], Amazon S3[50], Dropbox [29],

[50], Google Drive [27], [50], Microsoft Skydrive [28], Amazon Cloud Drive[63], BitTorrent Sync[64], SugarSync[65], Ubuntu One[26], huBic[66], Mega[67], Syncany [24] as well as different mobile cloud apps [15], [68], [69]. Quick and Choo [27]–[29] also identified that data erasing tools such as Eraser and CCleaner could not completely remove the data remnants from Dropbox, Google Drive, and Microsoft SkyDrive. From the literature, there is currently no work that focuses on forensic investigation of CloudMe SaaS cloud – a gap we aim to cover in this research.

### III. RESEARCH METHODOLOGY

We adopted the research methodology of Quick and Choo [27]–[29] and Teing et al. [25], [53], [70] in the design of our experiments. The first step of the experiment was to setup the test environments for the desktop and mobile clients. The former consisted of three Virtual Machines (VMs) with following configurations:

- Windows 8.1 Professional (Service Pack 1, 64-bit, build 9600) with 2GB RAM and 20GB hard drive.
- Ubuntu 14.04.1 LTS with 1GB RAM and 20GB hard disk.
- Mac OS X Mavericks 10.9.5 with 1GB RAM and 60GB hard drive.

The VMs were hosted using VMware Fusion Professional version 7.0.0 (2103067) on a Macbook Pro running Mac OS X Mavericks 10.9.5, with a 2.6GHz Intel Core i7 processor and 16GB of RAM. As explained by Quick and Choo [27]–[29], using physical hardware to undertake setup, erasing, copying, and re-installing an application would have been an onerous exercise. The client mobile devices comprised a factory restored iPhone 4 running iOS 7.1.2 and an HTC One X running Android KitKat 4.4.4. The mobile devices were jailbroken/rooted with 'Pangu8 Version 1.1' and 'Odin3 Version 185' (respectively) to enable root access. The $3111^{th}$ email messages of the Berkeley Enron email dataset (downloaded from *http://bailando.sims.berkeley.edu/enron_email.html*) were used to create a set of sample files which are saved in .RTF, .TXT, .DOCX, .JPG (print screen), .ZIP, and .PDF formats, providing dataset for this research experiment. Similar to previous studies [25], [28], [65], [71], we conducted a predefined set of experiments namely installation and uninstallation of the CloudMe client applications as well as uploading, downloading, viewing, deleting, unsyncing, sharing, and inactivating the sync files/folders to simulate various real world scenarios of using the CloudMe desktop and mobile client applications as well as web application using the Google Chrome client for Windows version 51.0.2704.103m. Before each experiment, we made a base snapshot of each VM workstation to serve as a control case. After each experiment, we created a snapshot of the VM workstations and took a bit-stream (dd) image of the virtual memory (.VMEM file) and a forensic copy of the virtual disk (.VMDK file) of the latter in Encase Evidence (E01) format. The decision to instantiate the physical memory dumps and hard disks with the virtual disk and memory files was taken to prevent the datasets from being modified because of using memory/image acquisition tools [27]–[29]. As for the mobile clients, we made binary images using 'dd' over SSH/ADB Shell.

In addition, we set up a forensic workstation with the tools in Table I to analyse collected evidences. We filtered collected data that matched the terms 'cloudme', 'xcerion, and 'Enron3111′ in the forensic images. These included SQLite database files, PLIST files, prefetch files, event logs, shortcuts, thumbnail cache, $MFT, $LogFile, $UsnJrnl, as well as web browser files (e.g., in *%AppData%\Local\Google*, *%AppData%\Local\Microsoft\Windows\WebCache*, *%AppData%\Roaming\Mozilla*, *%AppData%\Local\Microsoft\Windows\Temporary Files\index.dat*). Memory images were analysed using Volatility, Photorec file carver, and HxD Hex Editor. For all the data collected, both MD5 and SHA1 hash values were calculated and subsequently verified. All experiments were repeated thrice (at different dates) to ensure consistency of findings.

### IV. CLOUDME ANALYSIS ON DESKTOP CLIENTS

TABLE I
TOOLS PREPARED FOR SYNCANY FORENSICS

| Tools | Usage |
|---|---|
| FTK Imager Version 3.2.0.0 | To create a forensic image of the .VMDK files. |
| dd version 1.3.4-1 | To produce a bit-for-bit image of the .VMEM files. |
| Autopsy 3.1.1 | To parse the file system, produce directory listings, as well as extracting/analysing the files, Windows registry, swap file/partition, and unallocated space of the forensic images. |
| HxD Version 1.7.7.0 | To conduct keyword searches in the unstructured datasets. |
| Volatility 2.4 | To extract the running processes and network information from the physical memory dumps, dumping files from the memory space of the Syncany client applications, and detecting the memory space of a string (using the 'pslist', 'netstat'/'netscan', 'memdump' and 'yarascan' functions). |
| SQLite Browser Version 3.4.0 and SQLite Forensic Explorer version | To view the contents of SQLite database files. |
| Photorec 7.0 | To data carve the unstructured datasets. |
| File juicer 4.45 | To extract files from files. |
| BrowsingHistoryView v.1.60 | To analyse the web browsing history. |
| Nirsoft Web Browser Passview v1.58 | To recover the credential details stored in web browsers. |
| Nirsoft cache viewer, ChromeCacheView 1.56, MozillaCacheView 1.62, IECacheView 1.53 | To analyse the web browsing caches. |
| Thumbcacheviewer Version 1.0.2.7 | To examine the Windows thumbnail cache. |
| Windows Event Viewer Version 1.0 | To view the Windows event logs. |
| Console Version 10.10 (543) | To view log files. |
| Windows File Analyser 2.6.0.0 | To analyse the Windows prefetch and link files. |
| NTFS Log Tracker | To parse and analyse the $LogFile, $MFT, and $UsnJrnl New Technology File System (NTFS) files. |
| Plist Explorer v1.0 | To examine the contents of Apple Property List (PLIST) files. |

```sql
SELECT
  d.username AS 'Owner Name',
  a.folder_id AS 'Sync Folder ID',
  a.document_id AS 'Sync File ID',
  a.name AS 'Sync File Name',
  c.local_path AS 'Sync Folder Path',
  a.size AS 'File Size',
  a.modified_date AS 'Sync File Last Modified Date',
  c.created AS 'Folder Creation Time',
  c.last_run AS 'Folder Last Sync Time',
  b.deleted AS 'Folder is Deleted',
  c.inactivated AS 'Folder is inactivated',
  c.encrypted AS 'Folder is encrypted'
FROM syncfolder_document_table a
INNER JOIN syncfolder_folder_table b ON a.folder_id=b.child_folder_id
INNER JOIN syncfolder_table c ON c.folder_id=a.root_folder_id
INNER JOIN user_table d ON d.user_id=a.owner;
```

Fig. 2. The SQL query used to parse the synchronisation history from Cache.db.

The installation of the CloudMe desktop clients created the data directory at *%AppData%\Local\CloudMe, /home/<User Profile>/.local/share/CloudMe, /Users/<User Profile>/Library/Application Support/CloudMe* on the Windows, Ubuntu, and Mac OS desktop clients. The sync (download) folders were located in the OS' Documents directory, such as *%Users%\[User Profile]\Documents*, */home/[User Profile]/Documents*, */User/[User Profile]/Documents/* on the Windows, Ubuntu and Mac OS clients by default. Deleting the sync folders with the option "When delete folder in the cloud and all its content is selected." in the client applications, we observed that the sync folders remained locally but were removed completely from the server. In all scenarios, the data and download directories remained after the uninstallation of the client applications.

### A. Cache.db Database

The file synchronisation metadata and cloud transaction records could be predominantly located in the */%CloudMe%/cache.db* database (in the data directory). The tables of forensic interest are 'user_table', 'syncfolder_table', 'syncfolder_folder_table', and 'syncfolder_document_table'. The 'user_table' holds the property information of users which logged in from the desktop clients; 'syncfolder_table' maintains a list of metadata associated with the sync folder(s) added by or download to the local device; 'syncfolder_folder_table' keeps track of the tree structure for the sync folder(s); 'syncfolder_document_table' records the metadata associated with the synced files in the sync folder(s). Details of the table columns of forensic interest are presented in Table II.

| | Owner Name | Sync Folder ID | Sync File ID | Sync File Name | Sync Folder Path | File Size | Sync File Last Modified Date | Folder Creation Time | Folder Last Sync Time | Folder is Deleted | Folder is inactivated | Folder is encrypted |
|---|---|---|---|---|---|---|---|---|---|---|---|---|
| 24 | adamthomson | 562958569596136 | 4457417804 | Enron3111.jpg | C:/Users/anonymous/Documents/MacSyncFolder | 287937 | 2016-03-16 12:25:07 | 2016-03-15 22:06:55 | 2016-03-16 04:41:40 | NULL | false | false |
| 25 | adamthomson | 562958569596136 | 4457417805 | Enron3111.pdf | C:/Users/anonymous/Documents/MacSyncFolder | 31747 | 2016-03-16 12:25:10 | 2016-03-15 22:06:55 | 2016-03-16 04:41:40 | NULL | false | false |
| 26 | adamthomson | 562958569596136 | 4457417806 | Enron3111.rtf | C:/Users/anonymous/Documents/MacSyncFolder | 43360 | 2016-03-16 12:25:13 | 2016-03-15 22:06:55 | 2016-03-16 04:41:40 | NULL | false | false |
| 27 | adamthomson | 562958569596136 | 4457417807 | Enron3111.txt | C:/Users/anonymous/Documents/MacSyncFolder | 2734 | 2016-03-16 12:25:13 | 2016-03-15 22:06:55 | 2016-03-16 04:41:40 | NULL | false | false |
| 28 | adamthomson | 562958569596136 | 4457417808 | Enron3111.zip | C:/Users/anonymous/Documents/MacSyncFolder | 30967 | 2016-03-16 12:25:20 | 2016-03-15 22:06:55 | 2016-03-16 04:41:40 | NULL | false | false |

Fig. 1. An excerpt of the output of the SQLite query from the Cache.db database.

TABLE II
TABLES AND TABLE COLUMNS OF FORENSIC INTERESTS FROM CACHE.DB

| Table | Table Column | Relevance |
|---|---|---|
| user_table | user_id | A unique numerical user ID for the user(s) logged in from the local device. This ID could assist a practitioner in correlating any user-specific data that might have been obtained from other sources of evidence. |
| | username | Username provided by the user during registration. |
| | devicename | Device name provided by the user during registration. |
| | created | Holds the addition time of the user account(s) in datetime format. |
| syncfolder_table | owner | Owner's ID which correlates with the 'user_id' table column of the 'user_table' table. |
| | name | Folder name. |
| | local_path | Local directory path. |
| | cloud_path | Server's directory path. |
| | folder_id | A unique numeric folder ID for the sync folder(s). |
| | created | Folder creation date in datetime format |
| | last_run | Last sync time in datetime format. |
| | inactivated | Folder has been inactivated; 'true' if yes, 'false' if no. |
| | encrypted | Folder has been encrypted; 'true' if yes, 'false' if no. |
| Syncfolder_folder_table | name | Folder name which correlates with the 'name' table column of the 'syncfolder_table' table. |
| | root_folder_id | Folder ID for the root sync folder, which correlates with the 'folder_id' table column of the 'syncfolder_table' table. |
| | folder_id | Folder ID for the sync folder(s), including the folder ID for the subfolder(s). |
| | child_folder_id | A unique numeric folder ID for the subfolder(s) associated with the sync folder(s). The root folder retains its original folder ID unchanged. |
| | creation_date | Folder creation time in datetime format. |
| | deleted | Folder has been deleted; *NULL* if not deleted. |
| | owner | Owner's ID for the sync folder(s), which correlates with the 'user_id' table column of ' user_table' table. |
| syncfolder_document_table | owner | Owner's ID for the sync folder(s), which correlates with the 'user_id' table column of ' user_table' table. |
| | name | Folder name. |
| | root_folder_id | Folder ID for the root sync folder. |
| | folder_id | Folder ID for the sync folder(s), including the folder ID for the subfolder(s), which correlates with the 'child_folder_id' table column of the 'syncfolder_folder_table' table. |
| | document_id | A unique numeric document ID for the sync file(s). |
| | size | File size. |
| | modified_date | Last modified date in datetime format. |
| | checksum | MD5 checksum for the modified document. |
| | main_checksum | MD5 checksum for the original document. |

Fig. 1. shows the SQL query used to parse Cache.db and produce synchronization history shown in Fig. 2. Examination of the Windows registry revealed the username for the currently logged in user and the device name in *HKEY_USERS\<SID>\Software\CloudMe\Sync\startup\me* and *HKEY_USERS\<SID>\Software\CloudMe\Sync\<Username>\_xClientId* (respectively). The username can be a useful identifying information for the cache.db database's remnants i.e., locating copies of the 'user_table' data in physical memory dumps. The client ID is a unique 32-character alphanumeric string used to identify a CloudMe device, which can be used to correlate residual evidences.

In Ubuntu client both username and clientID were detected at */home/<User Profile>/.config/CloudMe/Sync.conf* file, by looking at values for entries 'me' (of the 'startup' property) and '_xClientId' (of the 'Username' property) respectively. In the Mac OSX client, Username and ClientID were detected in the 'startup.me' and '[Username].xClientId' properties of the */Users/<User Profile>/Library/Preferences/com.CloudMe.Sync.plist* file.

*B. Cloudme Log Files*

Log files play a vital role in an incident investigation [13]. The CloudMe log files are located in the 'logs' subdirectory and created daily and named as [Year-Month-Day].txt. Although the log file only recorded application errors, it was possible to identify the file synchronisation time alongside the sync path from the log entries such as "*2016-03-15 14:52:02: CloudMeUnthreaded: Request error: "/Users/alice/Documents/UbuntuShareFolder/UbuntuSubFolder/UbuntuSubFolder/Enron3111.docx" | "Error downloading https://os.cloudme.com/v1/users/12886417622/favorites/112112/webshare/UbuntuSubFolder/UbuntuSubFolder/Enron3111.docx - server replied: Not Found" Error number: 203*", "*2016-03-15 14:56:30: onSyncRequestFailed: "WindowsSubFolder/WindowsSubFolder/Enron3111.pdf" | Type: "Uploading" | Error: "7"*", "*2016-03-15 14:56:30: SYNC_FILE_NOT_FOUND—SYNC_FOLDER_NOT_FOUND: ( 0 ) "WindowsSubFolder/WindowsSubFolder/Enron3111.pdf" :*" and "*2016-03-15 14:51:52: addRemoveLocalFolder:Fail: "/home/suspectpc/UbuntuSyncFolder/UbuntuSubFolder"*". We could also recover the login time alongside the logged in username from the log

```
<fs:folder id='562958569591836' xmlns:xlink='http://www.w3.org/1999/xlink' xmlns:fs='http://xcerion.com/folders.xsd'>
    <fs:folder id='562958569603280' name='cloudme investigation'>
        <fs:tag type='update' value='718585' group='webshare' propagated='true' />
    </fs:folder>
    ...
</fs:folder>
```

Fig. 6. An excerpt of the folder listing metadata file.

```
<sync version="1.9.6" dName="WIN-KMM6MUN4701" clientId="{1cb0b304-6387-4813-88a8-1a2425fb1e06}">
    <syncfolder name="CloudMe" path="C:/Users/anonymous/Documents/CloudMe" hasSynchronized="true" upload="true" download="true"
        hotsync="true" cloudmefolder="true" favoritefolder="false" conflict="backup" cloudPath="xios://Documents/CloudMe" folderId="562958569591836"
        folderSyncMode="1" folderMode="2" foldertype="1" inactivated="false" lastSync="2016-03-15 12:47:25" />
    ...
</sync>
```

Fig. 5. An excerpt of the device-specific sync folder metadata file.

```
<?xml version="1.0"?>
- <favorites xmlns:os="http://a9.com/-/spec/opensearch/1.1/">
    <os:totalResults>3</os:totalResults>
    <os:startIndex>0</os:startIndex>
    <os:itemsPerPage>1000</os:itemsPerPage>
    <favorite document_id="0" folder_id="562958569591836" access="update" description="" password="digitalevidence" name="CloudMe" created="2016-03-16T04:41:34Z"
        webShareId="718585" sharingUserName="adamthomson" sharingUserId="12886417622" userId="12886417622" id="112118"/>
    <favorite document_id="0" folder_id="562958569596136" description="foldersharingfromMacOS " password="digitalevidence" name="MacSyncFolder" created="2016-03-17T04:57:49Z"
        webShareId="718534" sharingUserName="adamthomson" sharingUserId="12886417622" userId="12886417622" id="112124"/>
    <favorite document_id="0" folder_id="562958569596139" description="" password="digitalevidence" name="UbuntuSyncFolder" created="2016-03-15T14:43:17Z" webShareId="718533"
        sharingUserName="adamthomson" sharingUserId="12886417622" userId="12886417622" id="112112"/>
</favorites>
```

Fig. 4. The content of the extended=true&order=favoritename&count=1000&offset=0&_=1458191.xml document.

```
<?xml version="1.0"?>
- <webshares xmlns:os="http://a9.com/-/spec/opensearch/1.1/">
    <os:totalResults>6</os:totalResults>
    <os:startIndex>0</os:startIndex>
    <os:itemsPerPage>1000</os:itemsPerPage>
    - <webshare createdState="existing" updated="2016-03-16T04:41:12Z" created="2016-03-16T04:41:12Z" access="update" type="cloudme" password="digitalevidence" visibility="private"
        description="" name="CloudMe" userId="12886417622" id="718585">
        <folder name="CloudMe" id="562958569591836"/>
    </webshare>
    + <webshare createdState="existing" updated="2016-03-15T14:36:03Z" created="2016-03-15T14:36:03Z" access="update" type="" password="Digitalevidence" visibility="private"
        description="" name="IosSubFolder" userId="12886417622" id="718531">
    + <webshare createdState="existing" updated="2016-03-16T04:12:37Z" created="2016-03-16T04:12:37Z" access="update" type="" password="Digitalevidence" visibility="private"
        description="" name="IOSSyncFolder" userId="12886417622" id="718584">
    + <webshare createdState="existing" updated="2016-03-15T14:45:44Z" created="2016-03-15T14:45:44Z" access="read" type="cloudme" password="digitalevidence" visibility="private"
        description="foldersharingfromMacOS " name="MacSyncFolder" userId="12886417622" id="718534">
    + <webshare createdState="existing" updated="2016-03-15T14:42:39Z" created="2016-03-15T14:42:39Z" access="read" type="cloudme" password="digitalevidence" visibility="private"
        description="" name="UbuntuSyncFolder" userId="12886417622" id="718533">
    + <webshare createdState="existing" updated="2016-03-15T14:31:13Z" created="2016-03-15T14:31:13Z" access="read" type="cloudme" password="" visibility="private" description=""
        name="WindowsSyncFolder" userId="12886417622" id="718530">
</webshares>
```

Fig. 3. The content of the order=name&desc=false&count=1000&offset=0&resources=true&_=145.xml document.

entry "*2016-03-15 13:48:22: Logged in as: "adamthomson"*".

### C. Web Browser Artefacts

Web browsing activities history is a critical source of evidence [25], [27]–[29], [47]. Our analysis of the web browsing history found unique identifying URLs associated with the user actions. For example, when accessing a sync folder in the CloudMe web application, we observed following URLs:

- *https://www.cloudme.com/en#files:/Documents/<Folder name>*,
- *https://www.cloudme.com/en#files:/f:<Folder ID>*,
- *https://www.cloudme.com/en#sync:/f: <Folder ID>*,
- *https://www.cloudme.com/en#sync:/<Folder ID>*, and
- *https://www.cloudme.com/en#sync:/f:<Folder ID>, <Folder name>*.

Accessing or downloaded a sync file produced following URL:

- *https://www.cloudme.com/v1/documents/<Folder ID>/<Document ID>/1/<Filename>*.

When we accessed the folders shared with other users, we observed following URL:

- *https://www.cloudme.com/en#webshares:/<Folder name>*.

The download URL for the shared file could be discerned from:

- *https://www.cloudme.com/v1/documents/<Folder ID>/<Document ID>/1/<Filename>?dl=<Filename>*.

The web client's logout action generated following URL:

- *https://www.cloudme.com/en?r=1458192365602&logout=1*.

Rebuilding the web browsing caches produced the root directory for the web application at *www.cloudme.com/v1*. In particular, within the */%v1%/folders* directory we recovered a list of metadata files for the sync folders accessed by the user, which could be differentiated by the folder ID. Fig. 3 illustrates the metadata information associated with the sync folders; each of which creates a 'folder' subtag to house the folder ID and name, and a 'tag' subtag to hold the folder sharing information such as the webshare ID and folder sharing type i.e., in the 'group' property.

A search for the filenames of the sample files recovered files viewed on the web application in cache at

7*/%v1%/documents/<Folder ID>/<Document ID>/1/*. We also recovered thumbnails for the viewed files in */%v1%/documents/<Folder ID>/<Document ID>/<Thumbnail ID>*. Notice that the */%v1%/documents* directory will always contain at least one folder i.e., holding the metadata files associated with the sync devices at */%v1%/documents/<Folder ID>/<Document ID for device-specific metadata file>\1*. Fig. 4 shows that the device name and client ID can be detected from the 'dName' and 'clientId' properties of the 'sync' tag in the metadata file. Each sync folder creates a 'syncfolder' subtag to define the folder name, directory path, folder ID, last sync time, and information about whether the sync folder has been synchronised and if it is a favourite folder in the 'name', path', 'folderId,' lastSync', 'hasSynchronized', and 'favoriteFolder' properties respectively.

Another directory of interest within the 'v1' directory is the user-specific *%v1%\users\<User ID>* directory, which maintains a list of OpenSearch [72] description documents containing a wealth of folder metadata of forensic interest about the sync folders [73]. For example, the *%v1%/users/<User ID>/favorites/extended=true&order=favoritename&count=1000&offset=0&_=1458191.xml* document holds the OpenSearch description for the favourite folders. The metadata of interest recovered from this document include the folder IDs, folder names, folder sharing passwords, webshare IDs, as well as usernames and user IDs for the favourite folders in the 'folder_id', 'name', 'password', 'webShareId', sharingUserId', 'sharingUserName' properties of the sync folder/file-specific 'favorite' subtags (see Fig. 5). The *%v1%/users/<User ID>/webshares/order=name&desc=false&count=1000&offset=0&resources=true&_=145.xml* document defines the OpenSearch property of the shared folders/files, such as the update time, creation time, passwords, creators' IDs, webshare IDs in the 'updated', created', 'password', 'userId', and 'id' properties of the sync folder/file-specific 'webshare' subtags. The folder name and ID could be discerned from the 'name' and 'id' properties of the 'folder' subtag (see Fig. 6). Further details of the folder/file sharing could be located in the *%v1%/users/<User ID>/lifestream* document, such as the senders' user ID, senders' group ID, senders' username, receivers' user ID, receivers' group ID, receiver's username, favourite IDs (for favourite folders), and whether the sharing has been seen in the 'senderId', 'senderGroupId', 'senderName', 'receiverId', 'receiverGroupId', 'receiverName', 'parentFolder', and 'seen' properties in the 'event' subtags.

### D. Physical Memory Analysis

For all investigated client applications, analysis of the physical memory dumps using the 'pslist' function of Volatility recovered the process name, process identifier (PID), parent process identifiers (PPID) as well as process initiation time. We determined that the CloudMe process could be differentiated using the process names 'CloudMe.exe', 'cloudme-sync' and 'CloudMe' on the Windows, Ubuntu and Mac OS clients respectively.

Undertaking data carving of the memory image of the CloudMe process determined that the files of forensic interest such as cache.db, sync.config, and CloudMe logs can be recovered. When CloudMe was accessed using the web client, we could recover copies of the OpenSearch description documents containing the folder sharing passwords from the web browser's memory space intact. Unsurprisingly, we also managed to recover copies of the database, configuration, and log files in plain text. For the cache database, a search for the username for the user could locate the data [74] of the 'user_table', which holds the user ID in the row ID variant field of the cell header section [74] in hex format. Once the user ID is identified, a practitioner may locate the file offsets contained between the cell data section of the 'syncfolder_document_table', 'syncfolder_folder_table' and 'syncfolder_table' tables, and work backwards to read the header field type variants [74] to recover the remaining data fields.

## V. CLOUDME ANALYSIS ON MOBILE CLIENTS

Our examinations of the CloudMe mobile clients determined that the data directory is located in */private/var/mobile/Applications/<Universally Unique Identifier (UUID) for the CloudMe iOS app>/* and */data/data/com.excerion.android* on the iOS and Android clients. Although the mobile clients did not keep a copy of the sync folders from the user's account (like as the desktop clients), it was possible to recover copies of the viewed files from *%<Universally Unique Identifier (UUID) for the CloudMe iOS app>%/Documents/persistentCache/* and */storage/sdcard0/Android/data/com.xcerion.android/cache/files/Downloads/* of the iOS and Android clients by default.

### A. com.xcerion.icloud.iphone.plist and user_data.xml Files

A closer examination of the files in the directory listings located the username and password in plaintext in the 'username' and 'password' properties of the *%<Universally Unique Identifier (UUID) for the CloudMe iOS app>%/Library/Preferences/com.xcerion.icloud.iphone.plist* and *%com.excerion.android%/shared_prefs/user_data.xml* files. The former also held the last upload time in datetime format in the '<username_LastUploadTime>' property.

### B. db.sdb Database

Analysis of the Android client revealed the cache database at */storage/sdcard0/Android/data/com.xcerion.android/cache/db.sdb*. The tables of interest with the cache database are 'files' and 'folders'. The 'files' table maintains a list of metadata of the sync files viewed by the user, while the 'folders' table holds the metadata of the sync folders associated with the user's account. db.sdb Database shows the table fields of interest from the db.sdb database. We also proposed a SQL query to thread the table fields of interest from the tables to present the records in a forensically-friendly format as shown in Fig. 7.

## C. Cache.db Database

Further examination of the iOS client recovered copies of the responses for the web API queries in the *%<Universally Unique Identifier (UUID) for the CloudMe iOS app>%/Library/Caches/com.xcerion.icloud.iphone/nsurlcache/Cache.db* database. Specifically, we located the cached items in the 'receiver_data' table column of the cfurl_cache_receiver_data table in Binary Large OBject (BLOB) including metadata files and OpenSearch documents for the sync folders. Within the cfurl_cache_response table we located the corresponding URLs and timestamps in datetime format, in the "request_key" and "time_stamp" table columns, respectively. By threading the data fields using the SQL query "*SELECT cfurl_cache_receiver_data.receiver_data, cfurl_cache_response.request_key, cfurl_cache_response.time_stamp FROM cfurl_cache_receiver_data, cfurl_cache_response WHERE cfurl_cache_receiver_data.entry_ID=cfurl_cache_response.entry_ID*", it was possible to correlate the cached items with the URLs and timestamps.

## VI. CONCLUSION AND FUTURE WORK

In this paper, we examined the client residual artefacts left by CloudMe SaaS cloud as a backbone for big data storage. Our research included installing the client applications as well as uploading, downloading, deleting, sharing and activating/inactivating the sync folders/files using the client and web applications. We determined that a forensic practitioner investigating CloudMe cloud application should pay attention to the cache database, web caches, as well as log and configuration files as highlighted in TABLE IV. Unlike cloud applications such as Symform [25] and BitTorrent Sync [23], the CloudMe client applications did not create any identifying information (e.g., configuration file and cache folder) in the sync folders, and hence a practitioner cannot identify the sync directories from the directory listing. This also indicates that the cache database is critical source of evidence for the synchronisation metadata and cloud transaction records, and hence should not be overlooked.

TABLE III
TABLE FIELDS OF FORENSIC INTEREST FROM THE DB.SDB DATABASE.

| Table | Table column | Relevance |
|---|---|---|
| files | _id | A unique numerical user ID used to identify a CloudMe sync file. |
| | name | Filename for the sync file. |
| | folder_id | Folder ID for the folder housing the sync file. |
| | size | File size for the sync file. |
| | href | URL to the sync file. |
| | published | Sync file addition time in datetime format. |
| | updated | Last updated time of sync file in datetime format. |
| | owner | Owner's name of the sync file. |
| | Mime | Multipurpose Internet Mail Extensions (MIME) format of the sync file. |
| folders | Owner | Owner's name of the sync folder. |
| | Folder_id | A unique numerical user ID used to identify a CloudMe sync folder. |
| | Name | Folder's name. |
| | Parent | Folder's name for the parent folder. |
| | Is_root | Whether the sync folder is a root folder? |
| | Path | Original directory path for the sync folder. |

```sql
1  SELECT
2      a.owner AS 'Owner',
3      a.name AS 'Filename',
4      a.size AS 'File Size',
5      b.name AS 'Folder Name',
6      a.href AS 'URL',
7      a.published AS 'Published Time',
8      a.updated AS 'Last Updated Time',
9      a.mime AS 'File Type',
10     b.path AS 'Origin'
11 FROM files a
12 INNER JOIN folders b ON a.folder_id=b.folder_id
```

| | Owner | Filename | File Size | Folder Name | URL | Published Time | Last Updated Time | File Type | Origin |
|---|---|---|---|---|---|---|---|---|---|
| 1 | adamthomson | Enron3111.jpg | 689402 | AndroidSyncFolder | https://os.cloudme.com/v1/documents/5629585965596145/4457368187/1 | 2016-03-15T14:28:27Z | 2016-03-15T14:28:35Z | image/jpeg | xios://Documents/CloudMe/AndroidSyncFolder/ |
| 2 | adamthomson | Enron3111.docx | 78080 | AndroidSyncFolder | https://os.cloudme.com/v1/documents/5629585965596145/4457368325/1 | 2016-03-15T14:29:24Z | 2016-03-15T14:29:24Z | application/vnd.openxmlformats-officed... | xios://Documents/CloudMe/AndroidSyncFolder/ |
| 3 | adamthomson | cloudme_investigation.zip | 8939743 | cloudme investigation | https://os.cloudme.com/v1/documents/5629585969603280/4457426501/1 | 2016-03-16T11:53:52Z | 2016-03-16T11:53:52Z | application/zip | xios://Documents/CloudMe/cloudme investigation/ |

Fig. 7. The SQL query used to parse the file view history from the db.sdb database and the output.

Analysis of the mobile clients determined that the findings were not as conclusive in comparison with the desktop clients, and only the viewed files could be recovered. This indicated that the iOS and Android mobile clients are merely a UI for the web application. Our examination of the web browsing activities identified unique URLs that can aid in identification of the user actions made to the web application, such as login, logout, and accessing and downloading sync files/folders. Although the application layer was fully encrypted i.e., with the deployment of HTTPS, we were able to recover the root directory for the web application from the web browser's caches unencrypted, which included viewed files and metadata files and OpenSearch documents for the sync files/folders that contain the timestamp information and sharing passwords for the sync folders/files. However, a practitioner should note that the availability of the cached items depends on the API requests made to the web application and hence the artefacts may not be consistent across different occasions.

Our analysis of the physical memory captures revealed that the memory dumps may provide potential for alternative methods for recovering applications cache, logs, configuration files and other files of forensic interest. It was also possible to recover the folder sharing password from the web cache in plain text, but not for the login password. This suggested that a practitioner can only obtain the login password from the mobile clients, using WebBrowserPassView when manually saved in the web browsers, through an offline brute-force technique, or directly from the user. Nevertheless, a practitioner must keep in mind that memory changes frequently according to user activities and will be wiped as soon as the system is shut down. Hence, obtaining the memory snapshot of a compromised system as quickly as possible increases the likelihood of obtaining the encryption key before it is overwritten in memory.

TABLE IV
LOCATIONS OF FILES OF FORENSIC INTEREST FROM CLOUDME

| Content | Directory Paths |
|---|---|
| Database | • Cache.db database in *%AppData%\Local\CloudMe\*, */home/<User Profile>/.local/share/CloudMe/*, */Users/<User Profile>/Library/Application Support/CloudMe/*, and *%<Universally Unique Identifier (UUID) for the CloudMe iOS app>%/Library/Caches/com.xcerion.icloud.iphone/nsurlcache/* of the Windows, Ubuntu, Mac OS, and iOS clients.<br>• */storage/sdcard0/Android/data/com.xcerion.android/cache/db.sdb* on the Android client. |
| Log files | • [Year-Month-Day].txt in *%AppData%\Local\CloudMe\logs\*, */home/<User Profile>/.local/share/CloudMe/logs/*, */Users/<User Profile>/Library/Application Support/CloudMe/logs/*, and of the Windows, Ubuntu, and Mac OS clients. |
| Default download directory | • *%Users%\[User Profile]\Documents\*, */home/[User Profile]/Documents/*, */User/[User Profile]/Documents/*, *%<Universally Unique Identifier (UUID) for the CloudMe iOS app>%/Documents\persistentCache\*, and */storage/sdcard0/Android/data/com.xcerion.android/cache/files/Downloads/* on the Windows, Ubuntu, Mac OS, iOS, and Android clients. |
| Configuration files | *HKEY_USERS\<SID>\Software\CloudMe registry key*, */home/<User Profile>/.config/CloudMe/Sync.conf*, */Users/<User Profile>/Library/Preferences/com.CloudMe.Sync.plist*, *%<Universally Unique Identifier (UUID) for the CloudMe iOS app>%/Library/Preferences/com.xcerion.icloud.iphone.plist*, and *%com.excerion.android%/shared_prefs/user_data.xml* files on the Windows, Ubuntu, Mac OS, iOS, and Android clients. |
| Web caches | *www.cloudme.com/v1* directory of the CloudMe root directory for the web application. |

Future work would include extending this study to cloud storage services to have an up-to-date understanding of the big data artefacts from different cloud deployment models, which could lay the foundation for the development of data reduction techniques (e.g., data mining and intelligence analysis) for these technologies [2], [4], [75], [76].

[51] M. Scanlon, J. Farina, and M.-T. Kechadi, "BitTorrent Sync: Network Investigation Methodology," Sep. 2014.

[52] M. Scanlon, J. Farina, N. A. L. Khac, and T. Kechadi, "Leveraging Decentralization to Extend the Digital Evidence Acquisition Window: Case Study on BitTorrent Sync," *ArXiv14098486 Cs*, Sep. 2014.

[53] Y.-Y. Teing, D. Ali, K.-K. R. Choo, and M. Yang, "Forensic investigation of P2P cloud storage services and backbone for IoT networks: BitTorrent Sync as a case study," *Comput. Electr. Eng.*, no. 2016, 2016.

[54] Q. Do, B. Martini, and K.-K. R. Choo, "A Forensically Sound Adversary Model for Mobile Devices," *PLoS ONE*, vol. 10, no. 9, p. e0138449, Sep. 2015.

[55] Q. Do, B. Martini, and K.-K. R. Choo, "Is the data on your wearable device secure? An Android Wear smartwatch case study," *Softw. Pract. Exp.*, p. n/a–n/a, Jan. 2016.

[56] N. H. Ab Rahman, N. D. W. Cahyani, and K.-K. R. Choo, "Cloud incident handling and forensic-by-design: cloud storage as a case study," *Concurr. Comput. Pract. Exp.*, Jan. 2016.

[57] R. Marty, "Cloud Application Logging for Forensics," in *Proceedings of the 2011 ACM Symposium on Applied Computing*, New York, NY, USA, 2011, pp. 178–184.

[58] C. Shields, O. Frieder, and M. Maloof, "A system for the proactive, continuous, and efficient collection of digital forensic evidence," *Digit. Investig.*, vol. 8, Supplement, pp. S3–S13, Aug. 2011.

[59] S. Zawoad and R. Hasan, "Cloud Forensics: A Meta-Study of Challenges, Approaches, and Open Problems," *ArXiv13026312 Cs*, Feb. 2013.

[60] J. Dykstra and A. T. Sherman, "Design and implementation of FROST: Digital forensic tools for the OpenStack cloud computing platform," *Digit. Investig.*, vol. 10, pp. S87–S95, Aug. 2013.

[61] T. Gebhardt and H. P. Reiser, "Network Forensics for Cloud Computing," in *Distributed Applications and Interoperable Systems*, J. Dowling and F. Taïani, Eds. Springer Berlin Heidelberg, 2013, pp. 29–42.

[62] D. Quick and K.-K. R. Choo, "Forensic collection of cloud storage data: Does the act of collection result in changes to the data or its metadata?," *Digit. Investig.*, vol. 10, no. 3, pp. 266–277, Oct. 2013.

[63] Y.-Y. Teing, D. Ali, K.-K. R. Choo, M. Conti, and T. Dargahi, "Forensic Investigation of Cooperative Storage Cloud Service: Symform as a Case Study," *J. Forensics Sci.*, vol. [In Press], pp. 1–14, 2016.

[64] J. S. Hale, "Amazon Cloud Drive forensic analysis," *Digit. Investig.*, vol. 10, no. 3, pp. 259–265, Oct. 2013.

[65] J. Farina, M. Scanlon, and M.-T. Kechadi, "BitTorrent Sync: First Impressions and Digital Forensic Implications," *Digit. Investig.*, vol. 11, Supplement 1, pp. S77–S86, May 2014.

[66] M. Shariati, A. Dehghantanha, and K.-K. R. Choo, "SugarSync forensic analysis," *Aust. J. Forensic Sci.*, vol. 48, no. 1, pp. 1–23, Apr. 2015.

[67] B. Blakeley, C. Cooney, A. Dehghantanha, and R. Aspin, "Cloud Storage Forensic: hubiC as a Case-Study," in *2015 IEEE 7th International Conference on Cloud Computing Technology and Science (CloudCom)*, 2015, pp. 536–541.

[68] F. Daryabar, A. Dehghantanha, and K.-K. R. Choo, "Cloud storage forensics: MEGA as a case study," *Aust. J. Forensic Sci.*, pp. 1–14, Apr. 2016.

[69] F. Daryabar, A. Dehghantanha, B. Eterovic-Soric, and K.-K. R. Choo, "Forensic investigation of OneDrive, Box, GoogleDrive and Dropbox applications on Android and iOS devices," *Aust. J. Forensic Sci.*, vol. 48, no. 6, pp. 1–28, Mar. 2016.

[70] B. Martini, Q. Do, and K.-K. R. Choo, "Chapter 15 - Mobile cloud forensics: An analysis of seven popular Android apps," in *The Cloud Security Ecosystem*, Boston: Syngress, 2015, pp. 309–345.

[71] Y.-Y. Teing, D. Ali, K.-K. R. Choo, M. Zaiton, M. T. Abdullah, and W.-C. Chai, "A Closer Look at Syncany Windows and Ubuntu Clients' Residual Artefacts," in *Proceedings of 9th International Conference on Security, Privacy and Anonymity in Computation, Communication and Storage (SpaCCS 2016)*, Zhangjiajie, China, 16-18 November.

[72] N. H. A. Rahman and K.-K. R. Choo, "A survey of information security incident handling in the cloud," *Comput. Secur.*, vol. 49, pp. 45–69, Dec. 2014.

[73] A9.com Inc., "OpenSearch 1.1 Draft 5," 2016. [Online]. Available: http://www.opensearch.org/Specifications/OpenSearch/1.1#The_.22OpenSearchDescription.22_element. [Accessed: 26-May-2016].

[74] CloudMe AB, "CloudMe Web Services," 2016. [Online]. Available: https://www.cloudme.com/en/api/webservices. [Accessed: 26-May-2016].

[75] SQLite, "Database File Format," 2016. [Online]. Available: https://www.sqlite.org/fileformat.html. [Accessed: 10-Nov-2016].

[76] D. Quick and K.-K. R. Choo, "Impacts of increasing volume of digital forensic data: A survey and future research challenges," *Digit. Investig.*, vol. 11, no. 4, pp. 273–294, Dec. 2014.

[77] A. Dehghantanha, "Mining the Social Web: Data Mining Facebook, Twitter, LinkedIn, Google+, Github, and More," *J. Inf. Priv. Secur.*, vol. 11, no. 2, pp. 137–138, Apr. 2015.

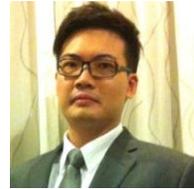

**Yee-Yang Teing** is a research fellow at the Putra University of Malaysia, and holds a Bachelor of Computer Forensics (First Class Honours). He is a Certified Ethical Hacker (CEH), Computer Hacking Forensic Investigator (CHFI) and Certified Security Analyst (ECSA). His research interests include cybercrime investigations, malware analysis and network security.

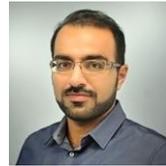

**Ali Dehghantanha** is a Marie-Curie International Incoming Fellow in Cyber Forensics and a fellow of the UK Higher Education Academy (HEA). He has served for many years in a variety of research and industrial positions. Other than Ph.D. in Cyber Security he holds many professional certificates such as GXPN, GREM, CISM, CISSP, and CCFP. He has served as an expert witness, cyber forensics analysts and malware researcher with leading players in Cyber-Security and E-Commerce.

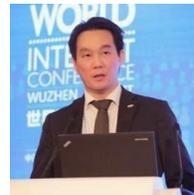

**Kim-Kwang Raymond Choo** (SM'15) received the Ph.D. in Information Security in 2006 from Queensland University of Technology, Australia. He currently holds the Cloud Technology Endowed Professorship at The University of Texas at San Antonio, and is an associate professor at University of South Australia, and a guest professor at China University of Geosciences, Wuhan. He is the recipient of various awards including ESORICS 2015 Best Paper Award, Winning Team of the Germany's University of Erlangen-Nuremberg (FAU) Digital Forensics Research Challenge 2015, and 2014 Highly Commended Award by the Australia New Zealand Policing Advisory Agency, Fulbright Scholarship in 2009, 2008 Australia Day Achievement Medallion, and British Computer Society's Wilkes Award in 2008. He is a Fellow of the Australian Computer Society.